\documentclass[12pt,preprint,iop]{emulateapj}

\usepackage{epsfig, natbib} % , deluxetable}
\usepackage{mathrsfs,amssymb}
\usepackage{graphicx,latexsym} % ,lscape}

\newcommand{\ang}{{\rm \AA}}

\newcommand{\etal}{{\rm et~al.\/}}

\newcommand{\mum}{\mbox{$\mu \rm m$}}

\slugcomment{Submitted to ApJ 2012 October 5, Accepted 2013 March 4}

\shorttitle{Far-IR Properties of Type 1 QSOs}
\shortauthors{Hanish \etal}

\begin{document}

\title{Far-Infrared Properties of Type 1 Quasars}

\author{D.J.\ Hanish\altaffilmark{1},
        H.I. Teplitz \altaffilmark{1},
        P.\ Capak\altaffilmark{1},
	V.\ Desai\altaffilmark{1},
	L.\ Armus \altaffilmark{1},
	C.\ Brinkworth \altaffilmark{1},
        T.\ Brooke\altaffilmark{1},
	J.\ Colbert \altaffilmark{1},
%L.\ Edwards \altaffilmark{2},
	D.\ Fadda\altaffilmark{1},
        D.\ Frayer\altaffilmark{2},
	M.\ Huynh \altaffilmark{3},
	M.\ Lacy \altaffilmark{4},
	E.\ Murphy \altaffilmark{5},
	A.\ Noriega-Crespo \altaffilmark{1},
	R.\ Paladini \altaffilmark{1},
	C.\ Scarlata \altaffilmark{6},
	S.\ Shenoy \altaffilmark{7},
	and\ the\ SAFIRES\ Team}

\altaffiltext{1}{Spitzer Science Center, California Institute of Technology, MC 220-6, 1200 E California Blvd., Pasadena, CA 91125, $hanish@ipac.caltech.edu$}
%\altaffiltext{2}{Astronomy Department, 260 Whitney Avenue, Yale University, New Haven, CT 06511}
\altaffiltext{2}{National Radio Astronomy Observatory, P.O. Box 2, Green Bank, WV 24944}
\altaffiltext{3}{International Centre for Radio Astronomy Research, M468, University of Western Australia, Crawley, WA 6009, Australia}
\altaffiltext{4}{National Radio Astronomy Observatory, 520 Edgemont Road, Charlottesville, VA 22903}
\altaffiltext{5}{The Observatories of the Carnegie Institution for Science, CA 91101}
\altaffiltext{6}{Minnesota Institute for Astrophysics, Schhol of Physics and Astronomy, University of Minnesota, Minneapolis, MN 55455}
\altaffiltext{7}{Space Science Division, NASA Ames Research Center, M/S 245-6, Moffett Field, CA 94035}

%==============================================================================
\begin{abstract}

We use the $Spitzer\ Space\ Telescope$ Enhanced Imaging Products (SEIP) and the Spitzer Archival Far-InfraRed Extragalactic Survey (SAFIRES) to study the spectral energy distributions of spectroscopically confirmed type 1 quasars selected from the Sloan Digital Sky Survey (SDSS).  By combining the Spitzer and SDSS data with the 2-Micron All Sky Survey (2MASS) we are able to construct a statistically robust rest-frame 0.1 -- 100 \mum\ type 1 quasar template.  We find the quasar population is well-described by a single power-law SED at wavelengths less than 20 \mum, in good agreement with previous work.  However, at longer wavelengths we find a significant excess in infrared luminosity above an extrapolated power-law, along with significant object-to-object dispersion in the SED.   The mean excess reaches a maximum of 0.8 dex at rest-frame wavelengths near 100 \mum.

\end{abstract}

%\dsp
%==============================================================================
\section{Introduction} \label{s:intro}

Quasars, or Quasi-Stellar Objects (QSOs), are thought to be super-massive black holes (M~$ > 10^{6}\ $M$_\sun$) actively accreting material \citep[e.g.][]{b:elvis94}.  The energy output by these objects may control the rate of star formation in galaxies \citep{b:hopkins06}, so understanding and characterizing these objects is key to galaxy evolution studies.  These luminous objects are characterized by broad emission lines in the rest-frame optical and ultraviolet and are believed to represent the peak accretion period for most black holes.  During this period black-hole accretion is expected to dominate the bolometric output of the galaxy \citep{b:hopkins06}, and so the SED template should represent the energy output of an actively accreting black hole.

The Spectral Energy Distribution (SED) of quasars is particularly important because it traces how much energy is output by the black hole at each wavelength, and is needed to separate the galaxy and quasar light when determining key properties such as the star formation rate or stellar mass of a galaxy \citep{b:fritz06}.  Unfortunately, spectral energy distributions in the mid- and far-infrared (FIR) wavelengths are poorly understood.  To date, infrared quasar SEDs have been limited mainly by small samples and survey biases.

In the absence of detailed spectrophotometry, SEDs can be generated by averaging broad-band photometric measurements, making appropriate extrapolations when data are absent \citep[e.g.][]{b:elvis94,b:bruzual03,b:richards06,b:shang11,b:wu11}.  In the mid-infrared (3 -- 100 \mum) the most comprehensive study to date is the work of \citet{b:polletta07}, which used an X-ray selected sample of 136 Active Galactic Nuclei (AGNs) detected at 24 \mum\ by the $Spitzer$ Multiband Imaging Photometer \citep[MIPS][]{b:rieke04}.  However, this work adopted a power-law extrapolation beyond 3 \mum\ due to a lack of robust mid-infrared data.  Later work by \citet{b:hatz08} attempted to extend these SEDs beyond 20 \mum\ using the Spitzer Wide-area InfraRed Extragalactic \citep[SWIRE,][]{b:lonsdale03} survey, but reached inconclusive results due to the small number of high signal-to-noise detections at 70 and 160 \mum.  More recent works \citep{b:hatz10,b:bonfield11,b:dai12,b:sajina12} have extended quasar SEDs out to observed wavelengths of 500 \mum, using small samples of $Herschel$ data.

In this paper, we will focus on the 20 -- 100 \mum\ SEDs of type 1 Quasars.  The goal of this work is to characterize this template SED and the intrinsic scatter around it which may indicate residual galaxy light and/or obscuration of the black hole.  To probe the 3 -- 100 \mum\ range we use the Spitzer Enhanced Imaging Products \citep[SEIP,][]{b:capak13} and the Spitzer Archival FIR Extragalactic Survey \citep[SAFIRES,][]{b:hanish13} which provide science quality mosaics and photometry for most Cryogenic Infrared Array Camera \citep[IRAC][]{b:fazio04} and MIPS data in the Spitzer Heritage Archive.  The Sloan Digital Sky Survey \citep[SDSS,][]{b:abazajian09} Data Release 7 provides a robust SED within the observed 0.3 -- 1 \mum\ range, and the Two Micron All-Sky Survey \citep[2MASS,][]{b:skrutskie06}, is used to fill in the 1 -- 3 \mum\ wavelength range.

The paper is organized as follows: Section~\ref{s:data} describes the data used by the SAFIRES template creation process, Section~\ref{s:SED} explains the methods used to generate our SED template, Section~\ref{s:results} presents the templates created through this process, and Section~\ref{s:conc} presents our conclusions.  We assume a $\Lambda$CDM cosmology ($\Omega_0 = 0.3, \Omega_\Lambda = 0.7$) with a Hubble constant of $H_0 = 70$ km s$^{-1}$ Mpc$^{-1}$.

%==============================================================================
\section{Data} \label{s:data}

\subsection{Parent Sample} \label{s:sample}

The process used to create an accurate quasar SED requires photometric data across a wide range of wavelengths.  We combine 15 wavelength bands ranging from the visible bands of the Sloan Digital Sky Survey to the far-infrared MIPS bands of the $Spitzer$ Space Telescope.  For inclusion in this work, sources were required to meet five criteria:
\begin{itemize}
\item{Inclusion in the SDSS Quasar Catalog, Data Release 7.  The optical limits associated with this catalog restricts this sample to type 1 quasars, as type 2 quasars lack the optical luminosity to be identified by the SDSS.  This is examined in more detail in \citet{b:reyes08}}
\item{A location in one of two large, well-studied extragalactic ($|$b$| \ge 20.0$ degrees) fields with deep $Spitzer$ coverage: the $Spitzer$ Deep Wide-Field Survey \citep[SDWFS,][]{b:ashby09} or the Lockman Hole \citep{b:lockman86}.}
\item{A SNR $\ge 10$ detection exceeding the minimum depth of coverage in at least one of the seven $Spitzer$ bands.  Only point sources were considered; sources showing extended emission were removed from the sample.}
\item{A redshift of $z \le 3.0$, provided by the SDSS Quasar Catalog.}
\item{An optical/near-infrared power-law SED slope of $\alpha \ge -1.2$, removing a small number of severely reddened type 1 QSOs from the sample.}
\end{itemize}

While the full SDSS Quasar Catalog \citep{b:schneider10} includes 105,873 objects, the first three criteria result in a sample of 328 type 1 QSOs possessing measurements in optical as well as mid-infrared wavelengths.  We further refine this sample of 328 sources with two additional criteria, to remove small numbers of outlying sources which would have interfered with our ability to reliably generate a characteristic type 1 template.  High-redshift sources tend to suffer from severe blending issues in the higher MIPS bands, and so seventeen of these 328 quasars are removed by our $z \le 3.0$ criterion.  Ten additional quasars were removed from the template generation algorithms by the fifth and final criterion, as they appeared to be highly dust-reddened type 1 QSOs as described by \citet{b:glikman12}.  While these ten quasars are still classified as type 1 QSOs, their SEDs are not typical of the general type 1 population.  As a result of our last two criteria, a total of 27 type 1 QSOs are removed from our SED template generation.

These five criteria result in a sample of 301 optically identified quasars with measurements in at least one $Spitzer$ infrared band.  As shown in Figure~\ref{f:zhist}, these quasars show a fairly even distribution of redshifts between 0.0 and 3.0, resulting in measurements across a smooth distribution of rest-frame wavelengths extending from the Lyman Break to beyond 100 \mum\ in the far-infrared (FIR).  This consistency is important for our template generation process, as a more restrictive redshift cutoff would result in gaps in the distribution of rest-frame wavelengths.  As the distribution of redshifts in this sample of 301 quasars matches closely that of the SDSS sample as a whole, we feel confident that our selection methods do not inherently bias towards or against quasars at certain redshifts.

{ % \ssp
\begin{figure*}[!t]
%\plotone{figure1.eps}
\centerline{\hbox{\includegraphics[width=5.5cm]{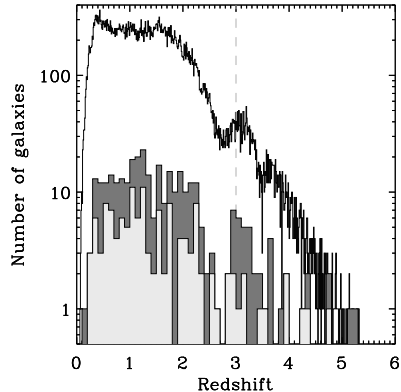}}}
\caption[Redshift Histogram]{The redshifts for the objects included in the SDSS and SEIP/SAFIRES samples.  The full SDSS catalog is shown in black, while the two lower histograms display all quasars located within the area observed by SAFIRES (dark grey), and those with coverage in either 70 \mum\ or 160 \mum\ (light grey).  Redshift bin sizes are 0.01 for the SDSS sample and 0.1 for SEIP/SAFIRES.}
\label{f:zhist}
\end{figure*}}

\subsection{Optical and Near-Infrared: SDSS and 2MASS}

To be included in this work, a quasar requires detection in the Sloan Digital Sky Survey, Data Release 7 \citep[SDSS,][]{b:abazajian09}.  The SDSS Quasar Catalog \citep{b:schneider10} identifies 105,873 distinct quasars within the full SDSS area, covering 9380 square degrees of sky.  Each potential quasar is confirmed through spectral examination by that work; no explicit emission line equivalent width limit was applied to rule out other types of objects, and only a conservative emission line FWHM limit ($> 1000$ km s$^{-1}$) was applied.  However, \citet{b:schneider10} use catalogs provided by previous SDSS works to explicitly remove type 2 quasars \citep{b:reyes08}, Seyfert galaxies \citep{b:hao05}, and BL Lac objects \citep{b:plotkin10}.  The result is a sample believed to consist only of type 1 quasars, an assumption we do not dispute.

Each SDSS quasar possesses measured luminosities in all five optical SDSS wavelength bands (u$^\prime$, g$^\prime$, r$^\prime$, i$^\prime$, z$^\prime$), with central wavelengths ranging from 350 to 850 nm.   The sample ranges from $0 < z < 6.5$ and all quasars are brighter than $m_{i^\prime} \le 19.1$, which effectively removes Type 2 QSOs from our potential sample.  As the SDSS Quasar Catalog only includes spectroscopically confirmed quasars, is unknown how many additional type 1 quasars could also fall into the SDSS-$Spitzer$ coverage region.  The sky-derived flux density limits within optical bands are consistently several orders of magnitude below the typical values for type 1 quasars, and so none of these quasars would lack SDSS data due to insufficient flux density.  The completeness of the SDSS sample is controlled entirely by the ability to identify quasars' spectra.

Each SDSS quasar is also compared to data from the Two Micron All-Sky Survey \citep[2MASS,][]{b:skrutskie06}, using both the 2MASS All-Sky Point Source Catalog (PSC) and the 2MASS Survey Point Source Reject Table.  While this survey covers the entire sky, its sensitivity is far lower than that of the SDSS, preventing us from acquiring near-infrared fluxes for most of the SDSS quasars.  As the available 2MASS catalogs list only point sources, no extended objects could be included in this sample; this criterion greatly reduces the potential for source confusion, and more closely matches the process used for our mid- and far-infrared bands.

% To reduce the possibility of erroneous measurements, we match the SDSS quasars directly to sources in the 2MASS All-Sky Point Source Catalog (PSC), as well as any sources within the 2MASS Survey Point Source Reject Table with confidence ratings above 90\%.  These catalogs used detections of discrete objects; sources had to either meet a threshold of SNR $\ge 7$ in a single band, or 5 in all three 2MASS bands.  No sources with extended flux were included, a criterion which also removed the majority of potentially confused sources.  Of the 311 quasars located within the area of coverage of $Spitzer$, 47 were located within a distance of 1 arcsecond of objects in the 2MASS PSC.  An additional 15 quasars correlated to objects in the Point Source Reject table above the 90\% threshold.  Additional quasars were located near sources in the rejection table at lower confidences, but we selected a conservative cutoff to ensure a reliable SED for those sources for which we quote fluxes.  While our rejection of extended sources or lower-confidence point sources reduces our total detections, only the brightest SDSS quasars correlated to point sources within the 2MASS bands, as the 5$\sigma$ limit on the PSC and Reject Table prevented many fainter quasars from being included.  This will be discussed in more detail in Section~\ref{s:SED}.

\subsection{Mid-Infrared: $Spitzer$ Enhanced Imaging Products (SEIP)}

The initial release of the Spitzer Enhanced Imaging Products \citep{b:capak13} improves upon existing datasets in two significant ways: the use of larger, deeper $Spitzer$ fields, by combining contiguous data from multiple observation requests (AORs), and use of additional indirect detection mechanisms to supplement the $SNR \ge 10$ direct measurements.  This mid-infrared archival data is now available, and contains sources in the four $Spitzer$ IRAC bands (at 3.6, 4.5, 5.8, and 8.0 \mum) as well as the 24 \mum\ band of MIPS.  The full SEIP will contain sources from approximately 1500 square degrees of sky, with the subset released at the present time spanning around 50\% of the final amount.  This subset contains over 20 million detected sources of all types, with an estimated 20 -- 30 million additional Galactic sources eventually to be added.  The initial release of the SEIP includes nine large, deep extragalactic fields; for this work, we select quasars from two of the largest contiguous fields, the Lockman Hole \citep{b:lockman86} and the field of the $Spitzer$ Deep Wide-Field Survey \citep[SDWFS,][]{b:ashby09}.

Objects are included in the SEIP Source List if they were directly detected at $\ge 10\sigma$ in any bands, and detections in multiple bands are treated as a single object if the detections in multiple bands have no more than 1 arcsecond separation between IRAC bands, or 3 arcseconds when comparing to the MIPS 24 \mum\ band.  While 301 SDSS-located type 1 quasars fall within the full area observed by $Spitzer$, only 222 -- 229 quasars were located within the area of adequate coverage depth within each IRAC band.  While their observed areas were equivalent, each IRAC band had a slightly different total due to depth of coverage differences or the effects of bad pixels.

\subsection{Far-Infrared: SAFIRES}

We refine our type 1 QSO template in far-infrared wavelengths using source catalogs from the SAFIRES project.  SAFIRES expands the SEIP to include measurements of point sources observed within the MIPS 70 and 160 \mum\ bands throughout the cryogenic phase of the $Spitzer$ mission.  Unlike SEIP, SAFIRES is limited to only include extragalactic ($|$b$| \ge 20.0$ degrees) fields.  With this limitation, and the lower total area covered by the MIPS 70 and 160 \mum\ bands, the full SAFIRES data set will span approximately 600 deg$^{2}$ of sky when completed.

The methods used to extract these FIR sources are extremely similar to those used for the MIPS 24 \mum\ measurements, as the software used to process the cryogenic $Spitzer$ data was constructed with these wavelengths included.  The primary differences between these two MIPS bands and the previous five $Spitzer$ bands involve use of the Germanium Reprocessing Tools (GeRT) to further refine data, along with an increase in the correlation distances to compensate for the larger PSFs and pixel scales of the two longer-wavelength MIPS bands.  Similar source identification mechanisms to those of the SEIP were used throughout the SAFIRES project.

\subsection{Detection Methods} \label{s:detect}

{ % \ssp
\begin{deluxetable*}{l | c | c | c c c c | c c c}
  \tablewidth{0pt}
  \tabletypesize{\small}
%  \rotate
  \tablecaption{SAFIRES data distributions \label{t:ourtotals}}
  \tablehead{\colhead{Number of} &
             \colhead{SDSS} &
             \colhead{2MASS} &
             \multicolumn{4}{c}{IRAC} &
	     \multicolumn{3}{c}{MIPS} \\
	     \colhead{quasars} &
             \colhead{(all bands)} &
             \colhead{(all bands)} &
%	     \colhead{$J$} &
%             \colhead{$H$} &
%             \colhead{$K_{S}$} &
             \colhead{3.6\mum} &
             \colhead{4.5\mum} &
             \colhead{5.8\mum} &
             \colhead{8.0\mum} &
             \colhead{24\mum} &
             \colhead{70\mum} &
             \colhead{160\mum} }
\startdata
Covered & 301 $^{\dag}$ & 301 & 219 & 221 & 215 & 221 & 224 & 164 & 172 \\
Direct & 301 $^{\ddag}$ & 59 $^{\ddag}$ & 169 & 173 & 162 & 171 & 151 & 11 & 0 \\
Indirect & N/A & N/A & 46 & 43 & 32 & 36 & 70 & 89 & 95 \\
Nondetections & N/A & N/A & 4 & 5 & 21 & 14 & 3 & 64 & 77
\enddata
\tablecomments{
\dag: Includes any SDSS-identified quasars within the areas observed by the $Spitzer$ Space Telescope within each band. No attempt is made to quantify the number of quasars within this area lacking SDSS detections.\\
\ddag: All objects in the provided catalogs possessed magnitudes for all bands on an instrument; these were treated as direct detections.}
\end{deluxetable*}}

As explained in Section~\ref{s:sample}, our core sample consisted of 301 type 1 quasars possessing $Spitzer$ observations in at least one infrared band.  Of these, 253 possess data within at least one MIPS band, necessary for the generation of our far-infrared template.  The number of sources located within the area observed by each band with sufficient depth of coverage, and the number of detections within those areas, are shown in Table~\ref{t:ourtotals}.  The less resolved bands nearly always overlap areas observed in more resolved bands; less than 10\% of the sources observed in MIPS 70\mum\ or 160 \mum\ lack coverage in MIPS 24 \mum, and almost all of the 24 \mum\ sources possess IRAC coverage as well.

We attempted to measure the magnitude of each source within the sample considered here, for each of the fifteen wavelength bands used in this work.  For each measurement, there were four possible outcomes:

\begin{itemize}
\item{``Direct'' detections.  These are sources identified by reliable, self-contained source detection algorithms, using no information from other bands.  For all $Spitzer$ bands, a signal-to-noise ratio of at least 10 is required, while the 2MASS catalogs required only a 7$\sigma$ detection (5$\sigma$ if detected within all three 2MASS bands).  The tools used to extract this information vary with instrument. with IRAC bands using Source Extractor \citep{b:bertin96} and MIPS bands using the APEX package within MOPEX \citep{b:makovoz05}.  The direct SDSS and 2MASS values are drawn from three provided source catalogs: the SDSS Quasar Catalog, the 2MASS All-Sky Point Source Catalog (PSC), and the 2MASS Survey Point Source Reject Table.  Objects in the Point Source Reject Table are only used if they possessed confidence ratings above 90\%, in order to ensure reliable SED generation.}

\item{``Indirect'' detections.  These consist of $Spitzer$ aperture flux measurements centered on positions provided by the SDSS Quasar Catalog, instead of using positions measured directly in the band in question.  As this process is positionally less accurate than the direct detections, larger apertures are used to ensure complete measurement of objects.  This results in an increase in quoted noise levels, and so we require only SNR $\ge 3$.  While the SDSS Quasar Catalog uses a similar method to add SNR $\ge 2$ 2MASS measurements to their sources, we use no analogous system for our own 2MASS measurements, and there exists no similar method for the SDSS catalog.}

\item{Nondetections.  As the SDSS Quasar Catalog formed the basis for the sample considered here, all quasars are detected in SDSS.  There are only a handful of true nondetections in any of the four IRAC bands, and only two quasars possess adequate 24 \mum\ MIPS coverage depth but lack successful detections.  While the 2MASS catalogs lack measurements for most of the SDSS sources, the majority of sources in are nondetections in these bands due to the prohibitive magnitude limits of the 2MASS projects; these magnitude limits will be discussed in more detail in Section~\ref{s:SED}.  Nondetections form a significant part of this sample in the MIPS bands, for which our provided data table gives 3$\sigma$ upper limits, although we make more realistic assumptions when generating SED templates.}

\item{Rejections.  Sources can be rejected from this sample for a number of technical or statistical reasons, even if there exist image data for those regions.  These reasons include extended features (common in the MIPS bands due to blending issues with nearby sources), cosmic rays, edge effects, and bad pixels.  The most common reason for rejection was lack of coverage, as the spatial coverage of the two far-IR MIPS bands were smaller than those of the IRAC bands, where less than half of the Lockman Hole region was observed in at least one of the two far-infrared bands.  A less common issue is that of insufficient coverage, with individual measurements for an object ignored if the coverage in that band's composite image fell below a nonzero band-specific threshold.  While most common in the $Spitzer$ MIPS bands, this last effect is most easily seen in the IRAC bands; all IRAC-covered objects possessed data for all four bands, but variations in depth of coverage occasionally led to the rejection of some IRAC bands for a given object, resulting in the differing numbers of IRAC-covered objects in Table~\ref{t:ourtotals}}.
\end{itemize}

As shown in Table~\ref{t:ourtotals}, the relative contributions of these categories vary significantly with wavelength band.  Indirect aperture measurements formed only a small part of this sample in the IRAC bands, and were in the minority for MIPS 24 \mum.  Conversely, very few 70 or 160 \mum\ sources had direct detections, with nearly all of the detections in these bands using the indirect method.

In total, 193 out of 301 sources within the SDWFS and Lockman Hole fields met the minimum coverage levels of our work in at least one FIR band, and these sources are essential for the construction of our far-infrared template.  While the areas covered by the MIPS 70 and 160 \mum\ bands are the same, the depth of coverage minima reduce the numbers of sources in both bands, with 167 possessing adequate 70 \mum\ coverage and 176 with 160 \mum, as many far-infrared fields lacked the depth necessary for reliable measurements.  The majority of these indirect far-infrared detections had directly detected counterparts in at least one of the five $Spitzer$ wavelength bands from SEIP.  In total, only 26 of the 70 \mum\ and 31 of the 160 \mum\ aperture detections had adequate depth of coverage in at least one mid-infrared SEIP bands while not possessing any direct detections, and another 21 quasars had sufficient depth in one of the two far-infrared bands but lacked coverage in MIPS 24 \mum\ or any of the IRAC bands.

Direct measurements were correlated purely by distance, with each source matched to the nearest qualifying source from other catalogs.  SDSS, 2MASS, and IRAC bands required a separation of no more than 1 arcsecond, while MIPS 24 \mum\ detections could vary by up to 3 arcseconds from their counterparts.  The 70 \mum\ MIPS sources were considered to match those of the best possible SDSS or SEIP counterpart within a match radius of 6 arcseconds, while the 160 \mum\ data required a separation of no more than 8 arcseconds.  These distances were selected by examining the correlations for several hundred bright point source objects within an initial sample of quasars, and selecting the typical distance at which no apparent mismatches occurred.  As these correlations suffer greatly from confusion issues in the higher MIPS bands, this distance does not scale linearly with the PSF radius of each band.  The conservative nature of this distance cutoff produces a correlated source list of high confidence, instead of a comprehensive source catalog.

However, while the correlation distances are chosen such that mismatches were minimized, the large size of the point spread function in the 70 and 160 \mum\ bands means that the potential for source confusion is much highter than in a more resolved mid-infrared band, as any aperture is likely to include flux contributions from a number of nearby objects as well.  The result is the distinct possibility of a single SAFIRES detection or aperture including flux contributions from multiple distinct sources, potentially increasing our estimated SED at far-infrared wavelengths.  Figure~\ref{f:stamps} illustrates the difficulties inherent in correlating far-infrared sources to more resolved bands.  As most 70 and 160 \mum\ ``sources'' are actually the combination of several distinct objects, their center positions are unlikely to fall within our correlation distances, shown as the black circle in each image.  This effect will be examined in more detail in Section~\ref{s:results}.  

{ % \ssp
\begin{figure*}[!t]
\plotone{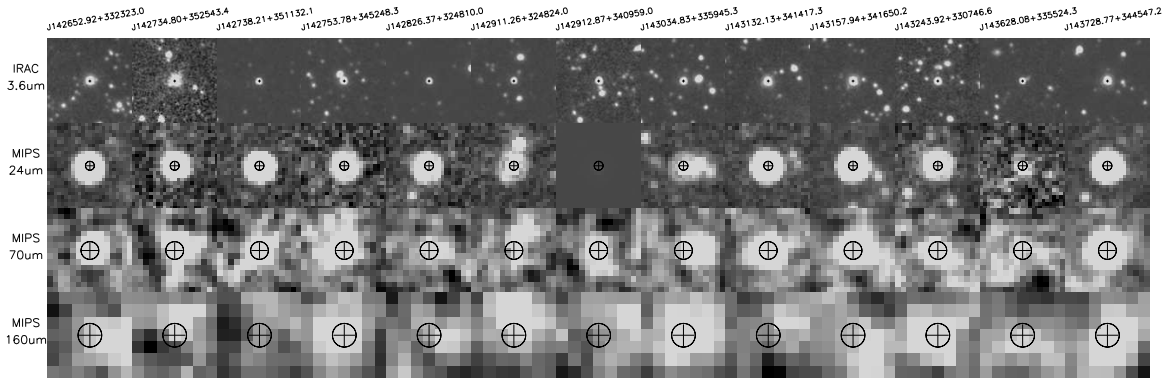}
%\centerline{\hbox{\includegraphics[width=15.5cm]{figure2.eps}}}
\caption[Cutouts]{Several quasars in the SDWFS field, shown in the IRAC 3.6 \mum, MIPS 24 \mum, MIPS 70 \mum, and MIPS 160 \mum\ bands. All images are 1 arcminute square, centered on the SDSS position for each quasar.  The overlay circles show the position of each quasar, and the maximum correlation distances allowed within each band.}
\label{f:stamps}
\end{figure*}}

% Detections are further divided into ``direct'' detections, where an object was found through source detection programs (Source Extractor \citep{b:bertin96} or the APEX package within MOPEX \citep{b:makovoz05}), and ``indirect'' detections, consisting of an aperture or PSF measurement method centered on a location drawn from a band with superior spatial resolution.

%==============================================================================
\section{Spectral Energy Distributions} \label{s:SED}

{ % \ssp
\begin{deluxetable*}{c | c | c}
  \tablewidth{0pt}
  \tabletypesize{\footnotesize}
  \tablecaption{SAFIRES photometry format \label{t:ourdata}}
  \tablehead{\colhead{Column} &
             \colhead{Name} &
             \colhead{Descriptions} }
\startdata
(1) & ID & Name from SDSS Quasar Catalog \\
(2) & RA & Right Ascension (J2000) \\
(3) & DEC & Declination (J2000) \\
(4) & Z & Spectroscopic redshift from SDSS \\
(5) & REGID & Region ID from SEIP; see \citet{b:capak13} \\
(6)--(10) & U\_MAG, G\_MAG, R\_MAG, I\_MAG, Z\_MAG & Magnitudes (AB) for quasars in the SDSS u$^\prime$, g$^\prime$, r$^\prime$, i$^\prime$, and z$^\prime$ bands \\
(11)--(13) & J\_MAG, H\_MAG, K\_MAG & Magnitudes (AB) for quasars in the 2MASS J, H, and K$_{S}$ bands \\
(14)--(17) & I1\_MAG, I2\_MAG, I3\_MAG, I4\_MAG & Magnitudes (AB) for quasars in the $Spitzer$ IRAC 3.6, 4.5, 5.6, and 8.0 \mum\ bands \\
(18)--(20) & M1\_MAG, M2\_MAG, M3\_MAG & Magnitudes (AB) for quasars in the $Spitzer$ MIPS 24, 70 and 160 \mum\ bands \\
(21)--(23) & J\_STAT, H\_STAT, K\_STAT & Detection status for 2MASS bands: Direct or Nondetection \\
(24)--(27) & I1\_STAT, I2\_STAT, I3\_STAT, I4\_STAT & Detection status for IRAC bands: Direct, Indirect, Nondectection, or Low Coverage \\
(28)--(30) & M1\_STAT, M2\_STAT, M3\_STAT & Detection status for MIPS bands: Direct, Indirect, Nondetection, or Low Coverage
\enddata
\tablecomments{The data table is available online for the 301 type 1 quasars used by the template generation algorithm.  For nondetections possessing adequate coverage, the quoted magnitudes are the 3$\sigma$ flux density limits at the appropriate location.}
\end{deluxetable*}}

By matching the $Spitzer$-observed sources from the SEIP and SAFIRES projects with their counterparts in the SDSS/2MASS quasar catalog, as detailed in Section~\ref{s:data}, we obtain a wide variety of data spanning up to fifteen bands with wavelengths ranging from 3500\ang\ to almost 200 \mum, with a robust population of rest-frame-adjusted effective wavelengths extending from the near-ultraviolet to around 100 \mum.  The measurements of each SAFIRES quasar for each available wavelength band are given in our online table, with a format described in Table~\ref{t:ourdata}.

% The table is too large to provide a sample here.

For the Lockman Hole, the areas containing sufficient depth of coverage in MIPS 70 and 160 \mum\ are less than half of those for IRAC, resulting in a somewhat lower number of potential quasars from those bands.  However, in the areas where seven $Spitzer$ bands were available, the detection rates for all bands remained comparable.  While IRAC and MIPS 24 \mum\ allowed for a large number of direct, SNR $\ge$ 10 detections, our indirect aperture method was sufficient to provide SNR $\ge$ 3 counterparts for these objects in the MIPS 70 and 160 \mum\ bands.  Likewise, the SDWFS field had excellent detection rates as the spatial coverages in all bands as the bands' respective areas overlapped substantially more, with over 90\% of the area in each region observed in all seven $Spitzer$ wavelength bands.  Within the 12.82 square degrees contained within SDWFS, the SDSS Quasar Catalog lists 81 quasars, of which 71 were directly detected in IRAC and the 24 \mum\ MIPS band, all 71 of which successfully acquired flux density values through our indirect method in both the 70 and 160 \mum\ MIPS bands.

The type 1 QSOs comprising this sample varied in luminosity by several orders of magnitude.  As most of these samples had incomplete data, with several bands lacking adequate depth or being removed for various technical reasons, it was necessary to normalize each object's flux densities to a common baseline before any sort of logic could combine the data points into a coherent template.  As we are trying to examine the shape of the far-infrared SED, we normalize each object based on its luminosities at rest-frame wavelengths between 0.2 and 1.0 \mum, the most well-studied part of the quasar SEDs.  Nearly every quasar (285 out of 301) possessed at least two measurements within this region, as every quasar in our sample had a complete set of five SDSS measurements, with many objects adding 2MASS detections as well.  More than half of our sample (166 objects) possess at least four observations within that rest-frame wavelength range, resulting in a robust normalization process.  All fluxes corresponding to rest-frame wavelengths within those limits were compared to the Polletta QSO1 template value at the appropriate wavelength, to derive a band-specific deviation from the QSO1 template; each object's fluxes were then normalized by the mean deviation for all bands within that rest-frame window.  This process effectively normalizes each object's SED by its optical and near-infrared intensities without penalizing objects lacking depth in individual bands, which is essential given the wide variation in spatial coverage for the various bands.

We used these normalized luminosities to generate a single SED template, with a method based on that of several earlier works \citep{b:richards06,b:polletta07,b:hatz08}.  Measurements are sorted into a series of rest-frame wavelength bins, which begin with a width of 0.2 dex.  This size was chosen to minimize the chance of an object possessing multiple rest-frame measurements within a single bin, a very real possibility when dealing with the optical SDSS bands.  We then use an iterative adjustment process to narrow the bins down further while keeping a minimum number of 100 rest-frame source measurements within each bin, with the final bins being as small as 0.03 dex.  The resulting distribution of binned data provides greater detail to the templates in a wavelength regime that previously used a simple power-law relationship, while keeping the number of detections within each bin at a reasonable value.  To minimize the impact of erroneous data points, we use the weighted medians within our wavelength bins to generate a final SED template, shown in yellow in Figure~\ref{f:SED}$a$.

%==============================================================================
\section{Results} \label{s:results}

\subsection{SED Template}

{ % \ssp
\begin{figure*}[!t]
\plotone{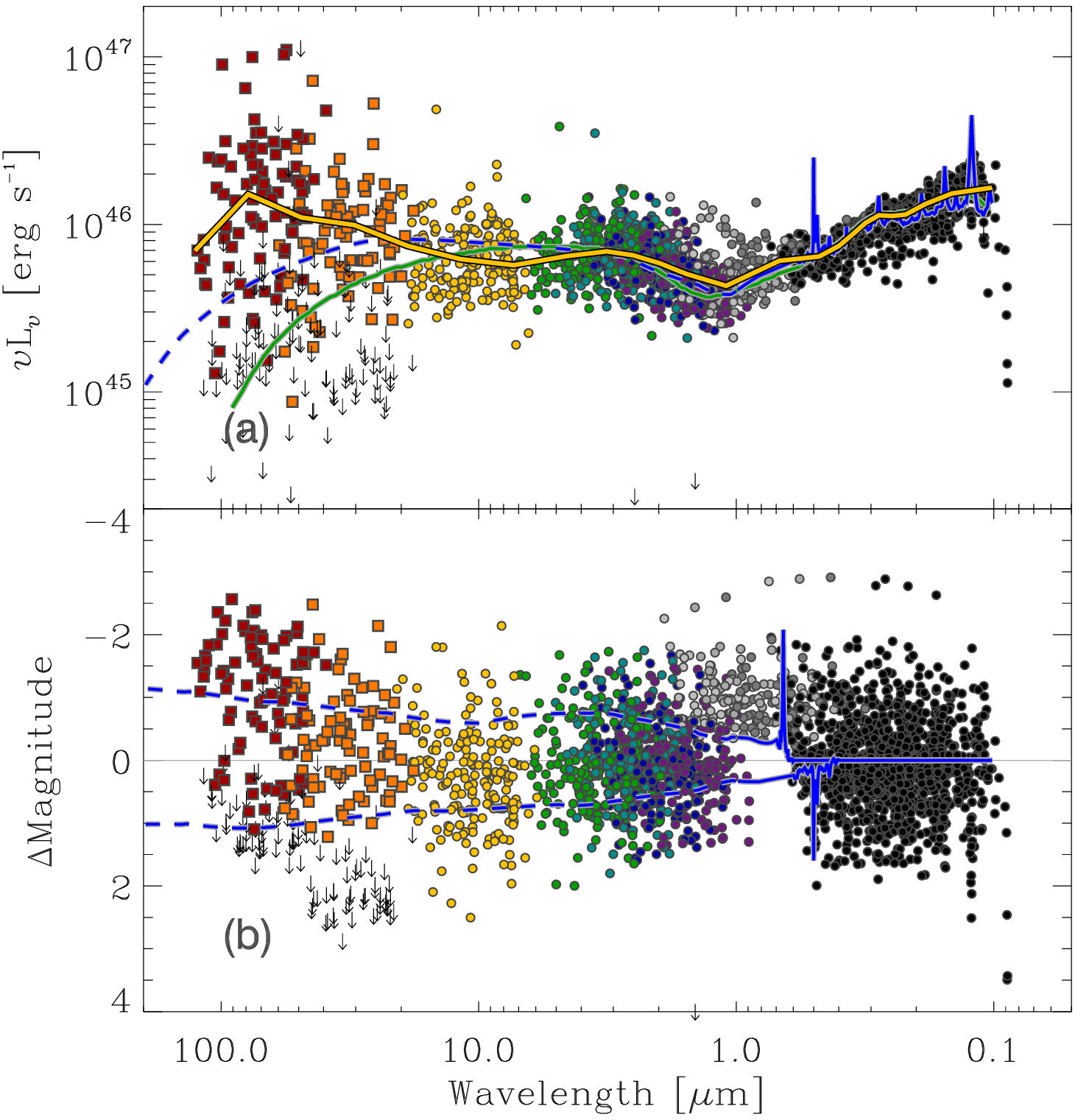}
\caption[Spectral energy distribution]{Spectral Energy Distributions of the SAFIRES quasar sample.\\
$(a)$: The rest-frame SED for the sample considered here, showing the normalized, binned median of detections (yellow, thick line), as well as normalized versions of the \citet{b:polletta07} QSO1 template (blue dashed) and \citet{b:richards06} template (green).  Colors of individual data points indicate the wavelength bands in which they were observed, with the SDSS bands in black, 2MASS bands in gray, and the seven $Spitzer$ bands using distinct colors (violet through red), with the two SAFIRES MIPS bands denoted with square symbols.  Arrows denote 3$\sigma$ flux limits for sources with no detections despite adequate depth of coverage.\\
$(b)$: The discrepancy between each non-normalized measurement and the QSO1 template at the corresponding rest-frame wavelength.  Blue lines denote the Polletta TQSO1 and BQSO1 quartile templates.}
\label{f:SED}
\end{figure*}}

We have created an SED template using the methods described in Section~\ref{s:SED}, for the SAFIRES-SDSS sample of 301 type 1 QSOs, 167 of which possess data in either the 70 or 160 \mum\ wavelength bands.  The results are shown in Figure~\ref{f:SED}$a$.  This template, using the binned median of detections, is shown in yellow, while the previous SED curves are shown in blue \citep[][QSO1]{b:polletta07} and green \citep{b:richards06}.  We have also confirmed the reliability of this template through the use of a Kaplan-Meier maximum-likelihood estimator \citep{b:feigelson85} to account for the far-infrared nondetections.  The Kaplan-Meier method allows for a robust estimation of the contributions of censored data, such as nondetections caused by our band-specific brightness limits, but the template remains dominated by the brightest sources and so the exact method used to estimate luminosities for our nondetections will have little impact on the resulting template.  Note that in Figure~\ref{f:SED}$a$, all object and SED luminosities are normalized based on their optical and near-infrared characteristics, as described in Section~\ref{s:SED}.  This counteracts the large inherent variation in our source measurements, with measured luminosity densities varying by up to five orders of magnitude across this sample; as the derived template's shape did not vary significantly between subsamples divided by luminosity or redshift, we feel justified in applying this normalization to our entire sample before calculating the median template.

The shape of the SAFIRES SED template matches well to the SDSS, 2MASS, IRAC, and MIPS 24 \mum\ data used by other studies, confirming the shape of earlier templates out to rest-frame wavelengths of approximately 10 \mum.  The width of the filters limits the effects of individual spectral lines; some earlier works avoided this issue by adding the spectral lines for a typical AGN to a binned median template, but we do not attempt to duplicate this process.  While the numbers of sources identified in the far-infrared MIPS bands are somewhat less than their counterparts in the lower-wavelength IRAC and SDSS bands, they are sufficient to add detail to a section of the standard SED templates that previous studies simply could not include.

As our quasar sample was drawn from the SDSS Quasar Catalog, with its inherent optical selection biases, no Type 2 QSOs, Seyfert galaxies, or BL Lac objects were included within any of these samples.  As explained in Section~\ref{s:sample}, ten additional dust-reddened type 1 QSOs were removed for having optical power-law fits far from typical of the remainder of the sample.  As a result, the template generated by SAFIRES can be seen as representative of the majority of type 1 QSOs.  As seen in Figure~\ref{f:SED}$a$, the SAFIRES rest-frame template generally matches the power-law relationship inherent to the previous templates within the mid-infrared regime.  However, we observe a significant deviation from these template at rest-frame wavelengths above 20 -- 30 \mum.  The template generated here appears to begin to deviate from the previous QSO1 template at approximately 20 \mum, increasing to an infrared excess of approximately 0.7 dex relative to the QSO1 template at rest-frame wavelengths of around 100 \mum.  As the \citet{b:richards06} template predicts an even lower far-infrared SED than \citet{b:polletta07}, our results deviate from it at a wavelength closer to 10 \mum, and the luminosity discrepancy near 100 \mum\ is more than one order of magnitude.  If the Richards template was truly representative of type 1 quasars, few of our quasars would have possessed far-infrared luminosities above the SAFIRES brightness limits.  As this turned out to not be the case in practice, we exclude the Richards template from further analysis and compare our results only to the \citet{b:polletta07} QSO1 template.

We wish to quantify the effect of the observed far-infrared excess on the total infrared luminosity of our quasar sample.  While the classical $L_{IR}$ extends from 8 \mum\ to 1000 \mum, the SAFIRES-generated SEDs only extend to approximately 150 \mum.  Over a more limited wavelength range, 8 -- 100 \mum, this template's infrared flux density exceeds that of its Polletta counterpart by a factor of 1.53, with nearly all of the discrepancy falling above a rest-frame wavelength of 80 \mum.  Determining the cumulative effect on IR luminosity out to 1000 \mum\ would require assumptions about the shape of the SED beyond the range of our observations.  If the quasar SED returns to the power-law relationship of earlier works above 200 \mum, then our observed discrepancy above 30 \mum\ would have a negligible effect on the total infrared value.  If this discrepancy is an indicator of a systematic underestimation of far-IR flux in earlier quasar templates, however, then the total disparity could easily dwarf the amount quoted above.

A similar far-infrared disparity has been observed in other works extending into the far-infrared, most recently \citet{b:dai12}, which used $Herschel$ SPIRE observations at 250, 350, and 500 \mum\ of a small sample of 32 quasars to estimate the mean quasar far-infrared SED.  In that study, the mean quasar deviates from the earlier power-law templates at rest-frame wavelengths above $\sim$10 \mum, by an average of 1.4 dex at a rest-frame wavelength of 90 \mum.  While our observed discrepancy is smaller, the disparity is most likely explained by statistical differences between the samples and methodology of the studies in question.  \citet{b:dai12} removes many fainter or possibly blended quasars, and divides its sample into ``FIR detected'' and ``FIR undetected'' subsamples, with the quoted 1.4 dex disparity being the average of the former.  For SAFIRES, most objects that would have fallen into the latter category or that would have been removed due to confusion issues can instead take advantage of our indirect methods to extract usable fluxes in the absence of a direct SNR $\ge 10$ detection.  As a result, we find the two studies to be in good agreement, confirming a significant far-infrared excess in quasars when compared to the templates of earlier studies.  

To examine these discrepancies in more detail, we compare the non-normalized quasar magnitudes to the QSO1 template in Figure~\ref{f:SED}$b$, to determine the point at which our median SAFIRES type 1 template significantly deviates from the power laws used by earlier works.  The earlier Polletta results include not only a mean type 1 QSO (QSO1) template, but ``top quartile'' (TQSO) and ``bottom quartile'' (BQSO) templates for the type 1 population as well.  While these quartiles are not true standard deviations, they can provide a similar function for our analysis.  Subtracting the QSO1 template from the SAFIRES data, without any normalization done on individual sources, we confirm that sources with data in these MIPS bands show a significant increase relative to the baseline template at far-IR rest-frame wavelengths, with medians falling well outside the curves corresponding to the QSO1 quartiles in the far-infrared regime.  Other than the 2MASS bands (shown in gray), which clearly suffer from their 5$\sigma$ limit on point sources, each band shows a comparable distribution of magnitudes up to a rest-frame wavelength of at least 20 \mum, but the far-infrared bands have considerably larger distributions of magnitudes.  The mean of this distribution does not significantly deviate from the quartile curves until a rest-frame wavelength of approximately 30 \mum, and the median discrepancy reaches its maximum value of approximately 2.0 magnitudes (0.8 dex) at around 80 \mum.  Few of our observed far-infrared data points (including only 10 observations in the 160 \mum\ MIPS band) fall below the previous Polletta QSO1 template, although approximately 40\% of quasars are not detected in each MIPS band.  These nondetections are not sufficient to counteract the excess in far-infrared quasar flux demonstrated by our observed quasars, and so our mean far-infrared luminosity appears to still fall well above the QSO1 template line.

Using Monte Carlo logic, we have also confirmed that the excess relative to the QSO1 template is not simply the statistical result of nondetections caused by the relatively high limiting fluxes in the far-infrared bands.  As a test, we assumed all of our quasar's SEDs had followed the QSO1 template's power-law relationship in the far-infrared, relative to their observed visible and near-infrared fluxes.  Under this assumption, more than half of our sources' far-infrared flux densities should have fallen below the 3$\sigma$ detection threshold.  However, the median of the detected sources would average only 0.1 -- 0.2 magnitudes higher than the QSO1 template, far less than the disparity we have observed.  This Monte Carlo simulation allows us to conclude that our observed increase in far-infrared flux density cannot be explained through purely statistical means.

This sample contains several inherent biases for redshifts, with high-redshift objects being significantly less likely to be successfully detected in the 70 or 160 \mum\ MIPS bands.  The completeness of the sample within each wavelength band is illustrated in Table~\ref{t:ourtotals}; while the majority of quasars possessed data in the IRAC band, the longer-wavelength MIPS bands had direct detections in a lower number of cases, generally losing the fainter high-redshift quasars within shallow $Spitzer$ fields.  This has the potential to raise our median template within the transition regions between bands, but this bias is difficult to quantify and does not appear to have a significant effect on our final templates.  No correction to magnitudes based on redshift was applied to the sample, as correctly applying the ``K correction'' requires assumptions about the shape of the spectrum of the object to which it is applied.  As our goal is to establish a template for the shape of that spectrum, these sort of assumptions are not feasible.  As previously mentioned, we have limited our SAFIRES template generation sample to include only those objects of redshift $z \le\ 3.0$, even though the input SDSS Quasar Catalog contained QSOs with redshifts of up to 6.0, as shown in Figure~\ref{f:zhist}.  While the regions possessing SAFIRES data were observed during several surveys of high-redshift or dusty quasars, the total population distribution remains comparable to that of the full SDSS sample.  We observed no significant changes in this sample when generating templates from only objects within a specific redshift range, such as only using objects at $z \le\ 1.0$.

\subsection{Confusion Bias}

{ % \ssp
\begin{figure*}[!t]
\plotone{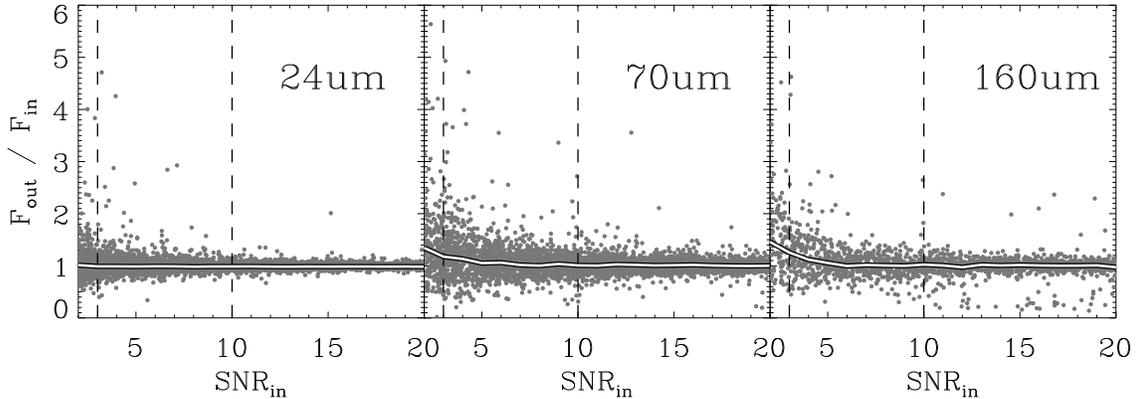}
\caption[SAFIRES SNR check]{The measured flux excess for an assortment of artificially created sources, relative to the intended signal-to-noise ratio of the false source.  The dashed lines denote the 10- and 3-$\sigma$ cutoffs used for direct detections and PSF/aperture measurements, respectively. The white lines denote binned medians within SNR bins of width 1.0 for each band. In none of the MIPS bands are the disparities between input and output flux large enough to explain our observed 0.8-dex discrepancy in the far-infrared SED.}
\label{f:SNRcheck}
\end{figure*}}

The most significant concern when processing far-infrared data is source confusion.  Unlike the other potential reasons for rejection, such as bad pixels or insufficient depth of coverage, this has the potential to create a significant and systematic discrepancy in the shape of the SED curve.  The primary cause of source confusion is tied to the sheer size of the Point Spread Function (PSF) in our less resolved bands.  For MIPS 160 \mum, the PSF has a width nearly seven times that of the MIPS 24 \mum\ band, affecting a region approximately 40 arcseconds in radius, while the MIPS 70 \mum\ PSF has a radius of around 17.5 arcseconds.  As a result, foreground objects that remain distinct in the more resolved bands, such as IRAC and the MIPS 24 \mum\ bands, may fall well inside the 70 \mum\ and 160 \mum\ apertures, and so be included in the measured flux for that quasar.

To determine whether this effect is sufficient to explain the noted flux discrepancy in the far-infrared, we create a number of false sources in existing images, using the shape of the PSF for each band, at random positions across a single 0.25 deg$^2$ area.  Object positions were only constrained to stay within the area possessing our minimum depths of coverage, approximately half of the total area; no constraints were used to limit proximity to true sources in each image.  These false objects were scaled to a wide range of intensities, and were designed to have randomly selected PSF signal-to-noise ratios ranging from 1.0 to 20.0, relative to the sky uncertainty of the entire image.  We repeat this process a number of times to ensure statistical confidence, creating a total of 20000 sources in MIPS 24 \mum within our single image.  Of this number, a total of 10461 false sources were located in areas possessing adequate depth of coverage, with the remaining 9539 sources removed from our analysis.  Similar methods were used in the other MIPS bands, with 4558 false sources placed within the same region in 70 \mum, and 2579 in the 160 \mum\ band.  The result is shown in Figure~\ref{f:SNRcheck}, comparing the ``true'' signal-to-noise ratios of our created false sources to the effective increases in flux caused by confusion.

The horizontal axis in Figure~\ref{f:SNRcheck} denotes the intensity of the false source relative to the image's background variation.  The quoted SNR for each object is directly proportional to its total flux, but any confusion would depend on the intensities of the potentially confused sources near the created object, and so correlate more closely with a local flux ratio, such as SNR, than with an absolute flux.  The vertical axis is simply the measured output flux relative to the known intensity of the created object.  High-SNR objects should, by definition, converge to a ratio of 1.0, while low-SNR objects would be expected to have higher ratios as nearby sources contribute proportionately more to the measured flux.  While there are several significant outliers, such as the small high-SNR population below the normal population in the 160 \mum\ band, most of these sources will not affect our actual results.  These outliers, were they real, would have been rejected from the SAFIRES sample for practical reasons, such as a PSF extending far outside the area of spatial coverage within a given band.  Additionally, the source measurement algorithm used for this check allows the measurement aperture to move towards local flux maxima within a limited radius, which would effectively increase the effects of source confusion in cases where a nearby object is brighter than the source to be measured.  As our quoted aperture fluxes do not allow for this movement, and instead are fixed at the position given by the SDSS Quasar Catalog, the bias quantities derived from this method will act as a maximum deviation.

While this method shows a notable excess when the faintest sources are created, the median does not significantly deviate from a 1.0 ratio for the vast majority of SAFIRES objects.  The median SAFIRES SNR for all detected 160 \mum\ sources is 8.54, despite every source in this band requiring only a 3$\sigma$ indirect measurement, instead of a 10$\sigma$ direct detection, and at that signal-to-noise ratio the median increase in flux due to confusion was less than 1\%.  Even for our faintest sources, those near the hard SNR $\ge 3$ limit required for fluxes added through our aperture system, the median ratio does not exceed 1.20 for 70 \mum\ data and 1.33 for 160 \mum.  As a result, we feel confident in our assumption that the noted SED discrepancy cannot be explained by any sort of systematic overmeasurement due to source confusion within the MIPS 70 and 160 \mum\ bands.

%==============================================================================
\section{Conclusions} \label{s:conc}

We have derived a type 1 QSO SED template using the algorithms described in Section~\ref{s:SED}, resulting in a robust template capable of modeling type 1 quasar spectra across the majority of infrared, optical, and ultraviolet wavelengths.  The quasar SED template discussed here confirms the findings of many recent works \citep[e.g.][]{b:hatz08,b:dai12}, in that the spectral energy distributions of quasars do not follow a single power law beyond wavelengths of 3 \mum, as was assumed in earlier works such as \citet{b:polletta07}.  Assuming that the sample used in this work is not biased towards any specific type of type 1 quasar, the SAFIRES template indicates a significantly higher flux above a rest-frame wavelength of 20 \mum\ when compared to previous templates, potentially resulting in a significantly increased total far-infrared luminosity.  Within this sample, it appears that while all type 1 quasars have similar SEDs in the visible and near-infrared regimes (barring the ten removed from our initial sample), their far-infrared characteristics can vary drastically.

\citet{b:hatz10} and \citet{b:bonfield11} explained the possible excesses in their far-infrared MIPS measurements through a combination of stellar emission and a cold dust component linked to bursts of star formation in an extended accretion disk.  Similar results were seen in the recent work of \citet{b:dai12}, where the projected dust temperatures needed to explain the observed far-infrared excess were typical associated with starburst galaxies.  Whether our observed discrepancy could be better modeled by further separating the type 1 QSO category, defined primarily by AGN characteristics, into multiple distinct subtypes with differing disk star formation properties is beyond the scope of this work.  It is possible that our observations could indicate a number of subclasses within the greater type 1 quasar category, with the various subclass' differences only becoming apparent at infrared wavelengths of more than 100 \mum\ as the other characteristics of the host galaxies begin to dominate.

Other far-infrared telescopes could be used to improve this work; many recent works \citep{b:hatz10,b:bonfield11,b:dai12,b:sajina12} use data from the $Herschel$ SPIRE instrument to examine quasars further into the far-infrared, allowing them to better characterize the downturn in quasar SEDs above a rest-frame wavelength of 100 \mum.  The $Herschel$ PACS instrument covers wavelengths similar to those of MIPS, albeit with greatly improved sensitivity and resolution, and so can perform an analysis similar to that shown in this work.  However, the primary constraints on the analysis in this work are those tied to the sample selection criteria in Section~\ref{s:sample}, such as the available far-IR coverage area and inclusion in the SDSS quasar catalog.  The spatial resolution and far-infrared depth of coverage of the instruments are less important than the total number of quasars observed.

As the Spitzer Enhanced Imaging Products and its SAFIRES offshoot continue to process and release archival data through the Spitzer Heritage Archive, the number of quasars with corresponding archival Spitzer data will rapidly increase.  The SEDs shown in this work will be updated with progressively larger numbers of sources as new data permits.  Additionally, the SDSS Quasar Catalog used in this project was based on SDSS Data Release 7; as the corresponding list for SDSS DR9 is expected to increase the number of verified quasars by almost 70\%, we would expect a comparable increase in the number of quasars included in our analysis.
\\
\acknowledgments

Support for this work was provided by NASA through contract NASA-ADAP NNX10AD52G.  This publication makes use of (1) raw data from the Spitzer Space Telescope, operated by the Jet Propulsion Laboratory/California Institute of Technology under NASA contract, (2) data products from the Two Micron All Sky Survey, a joint project of the University of Massachusetts and the Infrared Processing and Analysis Center/California Institute of Technology, funded by NASA and the NSF, and (3) the SDSS, managed by the Astrophysical Research Consortium for the Participating Institutions and funded by the Alfred P. Sloan Foundation, the Participating Institutions, and many others.

%==============================================================================


\begin{thebibliography}{}

%\bibitem[Atek \etal (2009)]{b:atek09}
%Atek, H., Kunth, D., Schaerer, D., Hayes, M., Deharveng, J.M., $\ddot{\rm O}$stlin, G., \&\ Mas-Hesse, J.M. 2009, A\&A, 506, L1

\bibitem[Abazajian \etal (2009)]{b:abazajian09}
Abazajian, K.N., \etal\ 2009, ApJS, 182, 543

\bibitem[Ashby \etal (2009)]{b:ashby09}
Ashby, M.L.N., \etal\ 2009, ApJ, 701, 428

\bibitem[Bertin \&\ Arnouts (1996)]{b:bertin96}
Bertin, E. \&\ Arnouts, S., 1996, A\&AS, 117, 393

\bibitem[Bonfield \etal (2011)]{b:bonfield11}
Bonfield, D.G., \etal\ 2011, MNRAS, 416, 13

\bibitem[Bruzual \&\ Charlot (2003)]{b:bruzual03}
Bruzual, A.G. \&\ Charlot, S., 2003, MNRAS, 344, 1000

\bibitem[Budav\'{a}ri \etal (2000)]{b:budavari00}
Budav\'{a}ri, T., Szalay, A.S., Connolly, A.J., Csabai, I., \&\ Dickinson, M. 2000, AJ, 120, 1588

\bibitem[Budav\'{a}ri \etal (2001)]{b:budavari01}
Budav\'{a}ri, T., \etal\ 2001, AJ, 122, 1163

\bibitem[Budav\'{a}ri \&\ Szalay (2008)]{b:budavari08}
Budav\'{a}ri, T., \&\ Szalay, A.S. 2008, ApJ, 679, 301

\bibitem[Capak \etal (in prep)]{b:capak13}
Capak, P., \etal, $in\ preparation$

\bibitem[Chiappetti \etal (2005)]{b:chiappetti05}
Chiappetti, L., \etal\ 2005, A\&A, 439, 413

%\bibitem[Cutri \etal (2003)]{b:cutri03}
%Cutri, R.M., \etal\ 2003, 

\bibitem[Dai \etal (2012)]{b:dai12}
Dai, Y.S., \etal\ 2012, ApJ, 753, 33

\bibitem[Elvis \etal (1994)]{b:elvis94}
Elvis, M., \etal\ 1994, ApJS, 95, 1

\bibitem[Fazio \etal (2004)]{b:fazio04}
Fazio, G. G., \etal\ 2004, ApJS, 154, 10

\bibitem[Feigelson \&\ Nelson (1985)]{b:feigelson85}
Feigelson, E.D., \&\ Nelson, P.I. 1985, ApJ, 293, 192

\bibitem[Fritz \etal (2006)]{b:fritz06}
Fritz, J., Franceschini, A., \&\ Hatziminaoglou, E., 2006, MNRAS, 366, 767

\bibitem[Glikman \etal (2012)]{b:glikman12}
Glikman, E., \etal\ 2012, ApJ, 757, 51

\bibitem[Hanish \etal (in prep)]{b:hanish13}
Hanish, D.J., \etal, $in\ preparation$

\bibitem[Hao \etal (2005)]{b:hao05}
Hao, L., \etal\ 2005, AJ, 129, 1783

\bibitem[Hatziminaoglou \etal (2008)]{b:hatz08}
Hatziminaoglou, E., \etal\ 2008, MNRAS, 386, 1252

\bibitem[Hatziminaoglou \etal (2010)]{b:hatz10}
Hatziminaoglou, E., \etal\ 2010, A\&A, 518, L33

\bibitem[Hopkins \etal (2006)]{b:hopkins06}
Hopkins, P.F., \etal\ 2006, ApJS, 163, 1

\bibitem[Lacy \etal (2004)]{b:lacy04}
Lacy, M., \etal\ 2004, ApJS, 154, 166

\bibitem[Lonsdale \etal (2003)]{b:lonsdale03}
Lonsdale, C.J., \etal\ 2003, PASP, 115, 897

\bibitem[Lockman \etal (1986)]{b:lockman86}
Lockman, F.J., Jahoda, K., \&\ McCammon, D. 1986, ApJ, 302, 432

\bibitem[Makovoz \&\ Marleau (2005)]{b:makovoz05}
Makovoz, D., \&\ Marleau, F.R. 2005, PASP, 117, 1113

\bibitem[Plotkin \etal (2010)]{b:plotkin10}
Plotkin, R.M., \etal\ 2010, AJ, 139, 390

\bibitem[Polletta \etal (2007)]{b:polletta07}
Polletta, M., \etal\ 2007, ApJ, 663, 81

\bibitem[Reyes \etal (2008)]{b:reyes08}
Reyes, R., \etal\ 2008, AJ, 136, 2373

\bibitem[Richards \etal (2006)]{b:richards06}
Richards, G.T., \etal\ 2006, ApJS, 166, 470

\bibitem[Rieke \etal (2004)]{b:rieke04}
Rieke, G.H., \etal\ 2004, ApJS, 154, 290

\bibitem[Sajina \etal (2012)]{b:sajina12}
Sajina, A., Yan, L., Fadda, D., Dasyra, K., \&\ Huynh, M. 2012, ApJ, 757, 13

%\bibitem[Schneider \etal (2007)]{b:schneider07}
%Schneider, D.P., \etal\ 2007, AJ, 134, 102

\bibitem[Schneider \etal (2010)]{b:schneider10}
Schneider, D.P., \etal\ 2010, AJ, 139, 2360

\bibitem[Shang \etal (2011)]{b:shang11}
Shang, Z., \etal\ 2011, ApJS, 196, 2

\bibitem[Skrutskie \etal (2006)]{b:skrutskie06}
Skrutskie, M.F., \etal\ 2006, AJ, 131, 1163

\bibitem[Wu \etal (2011)]{b:wu11}
Wu, Y., \etal\ 2011, ApJ, 734, 40

\end{thebibliography}
\end{document}